\documentclass[a4paper]{jpconf}
\usepackage{graphicx}
\usepackage{physics}
\usepackage{titlesec}
\usepackage{multirow}
\usepackage[sort&compress,square,comma,numbers]{natbib}
\usepackage{lipsum}
\usepackage{hyperref}
\usepackage{cleveref}
\usepackage{amsfonts}
\usepackage{xcolor}
\usepackage{subcaption}

\hypersetup{
    colorlinks = true,
    citecolor = magenta,
    urlcolor = blue,
}

\titlespacing*{\section}{0pt}{1.0ex}{1.0ex}
\titlespacing*{\subsection}{0pt}{1.0ex}{1.0ex}

\numberwithin{equation}{section}
\begin{document}

\title{Equivariant Graph Neural Networks for Charged Particle Tracking}

\author{%
Daniel Murnane$^1$, Savannah Thais$^2$, Ameya Thete$^3$
}

\address{$^1$ Scientific Data Division, Lawrence Berkeley National Laboratory, Berkeley, CA 94720, USA}
\address{$^2$ Data Science Institute, Columbia University, New York, NY 10027, USA}
\address{$^3$ Department of Physics, BITS, Pilani -- KK Birla Goa Campus, Zuarinagar, Goa 403726, India}

\ead{dtmurnane@lbl.gov}

\begin{abstract}
Graph neural networks (GNNs) have gained traction in high-energy physics (HEP) for their potential to improve accuracy and scalability. However, their resource-intensive nature and complex operations have motivated the development of symmetry-equivariant architectures. In this work, we introduce EuclidNet, a novel symmetry-equivariant GNN for charged particle tracking. EuclidNet leverages the graph representation of collision events and enforces rotational symmetry with respect to the detector's beamline axis, leading to a more efficient model. We benchmark EuclidNet against the state-of-the-art Interaction Network on the TrackML dataset, which simulates high-pileup conditions expected at the High-Luminosity Large Hadron Collider (HL-LHC). Our results show that EuclidNet achieves near-state-of-the-art performance at small model scales ($<1000$ parameters), outperforming the non-equivariant benchmarks. This study paves the way for future investigations into more resource-efficient GNN models for particle tracking in HEP experiments.
\end{abstract}

% Notes about the manuscript:
% Important: We need to stick to a 5 page limit (excluding references). 
% I'm sticking to American English spelling conventions. 

% Introduce the problem, provide a broad description of extant tracking methods in HEP and a 
% short description of our work to place it in context
\section{Introduction}
%\comment{
%\begin{itemize}
%    \item Mention that FPGAs could be a concrete application due to small memory, or symmetry constraint as an alternative to pruning + quantization
%\end{itemize}
%}

In recent years there has been a sharp increase in the use of graph neural networks (GNNs) for high-energy physics (HEP) analyses \cite{Thais:2022iok}. These studies have demonstrated that GNNs have the potential to deliver large improvements in accuracy, and can easily scale to sizable volumes of data. However, many of these architectures either involve a large number of parameters or complex graph operations and convolutions, which make GNNs resource-intensive and time-consuming to deploy. Many real-world datasets, including those from HEP experiments, display known symmetries, which can be used to construct alternatives to computationally expensive unconstrained architectures. By exploiting the inherent symmetry in a given problem, we can restrict the function space of neural networks to relevant candidates by enforcing equivariance with respect to transformations belonging a certain symmetry group. This approach has a two-fold benefit: incorporating equivariance introduces inductive biases into the neural network, and equivariant models may be more resource-efficient than their non-equivariant counterparts \cite{satorras2021n, Bogatskiy:2022hub, Murnane_2023}.

In this work, we introduce a new architecture of symmetry-equivariant GNNs for charged particle tracking. Using a graph representation of a collision event, we propose EuclidNet, which scalarizes input tracking features to enforce rotational symmetry. In particular, given the detector's symmetry around the beamline ($z$) axis, we develop a formulation of EuclidNet equivariant to the SO(2) rotation group. Benchmarked against the current state-of-the-art (SoTA) Interaction Network, EuclidNet achieves near-SoTA performance at small model sizes. We explore the out-of-distribution inference performance of EuclidNet and InteractionNet in order to provide an explanation for the behavior of model performance versus model size. As such, we reveal hints not only as to the upper ceiling on fully-equivariant models, but the symmetries learned by non-equivariant models. The code and models are publicly available at \url{https://github.com/ameya1101/equivariant-tracking}.

\vspace{1em}

% Introduce relevant text on equivariance and GNNs to complement the definition of EuclidNet.
% Needs a section on equivariance, GNNs, and conclude by integrating ideas from the previous two sections to define EuclidNet.
\section{Theory and Background}
In this section, we introduce the concept of symmetry-group equivariance and provide a short introduction to the theory of graph neural networks.

\subsection{Equivariance}
Formally, if $T_g: X \rightarrow X$ is a set of transformations on a vector space $X$ for an abstract symmetry group $g \in G$, a function $\phi: X \rightarrow Y$ is defined to be equivariant to $g$ if there exists an equivalent set of transformations on the output space $S_g: Y \rightarrow Y$ such that: 
\begin{equation}
    \phi(T_g(\vb{x})) = S_g\phi(\vb{x})
\end{equation}
A model is said to be equivariant to a group $G$ if it is composed of functions $\phi$ that are equivariant to $G$. In this work, we limit our discussion to rotational equivariance corresponding to the SO(2) group. As an example, let $\vb{x} = (\vb{x}_1, \vb{x}_2, \cdots, \vb{x}_M)$ be a set of $M$ points embedded in $n-$dimensional space, and $\phi(\vb{x}) = \vb{y} \in \mathbb{R}^{M \times n}$ be the transformed set of points. For an orthogonal rotation matrix $Q \in \mathbb{R}^{n \times n}$, $\phi$ is equivariant to rotations if $Q\vb{y} = \phi(Q\vb{x})$.

\subsection{Graph Neural Networks}
Consider a graph $\mathcal{G} = (V, E)$ with nodes $v_i \in V$ and edges $e_{ij} \in E$. A graph neural network is a permutation-invariant deep learning architecture that operates on graph-structured data \cite{Kipf:2016gmz}. A GNN commonly consists of multiple layers, with each layer performing the graph convolution operation, which is defined as \cite{gilmer2017neural}: 
\begin{align}
    \vb{m}_{ij} &= \phi_e(\vb{h}^{l}_{i}, \vb{h}^{l}_{j}, a_{ij})\\
    \vb{m}_i &= \sum_{j \in \mathcal{N}(i)} \vb{m}_{ij}\\
    \vb{h_{i}^{l + 1}} &= \phi_h(\vb{h}^{l}_{i}, \vb{m}_i)
\end{align}
where $\vb{h}^{l}_{i} \in \mathbb{R}^k $ is the $k$-th dimensional embedding of node $v_i$ at layer $l$. $a_{ij}$ are edge attributes. $ \mathcal{N}(i)$ is the set of neighbors of node $v_i$. Finally, $\phi_e$ and $\phi_h$ are edge and node operations which are approximated by multi-layer perceptrons (MLPs). $\vb{m}_{ij}$ are called messages that are passed between nodes $v_i$ and $v_j$. 

\vspace{1em}

\section{Network Architecture}
\begin{figure}[!ht]
    \centering
    \includegraphics[scale=0.2]{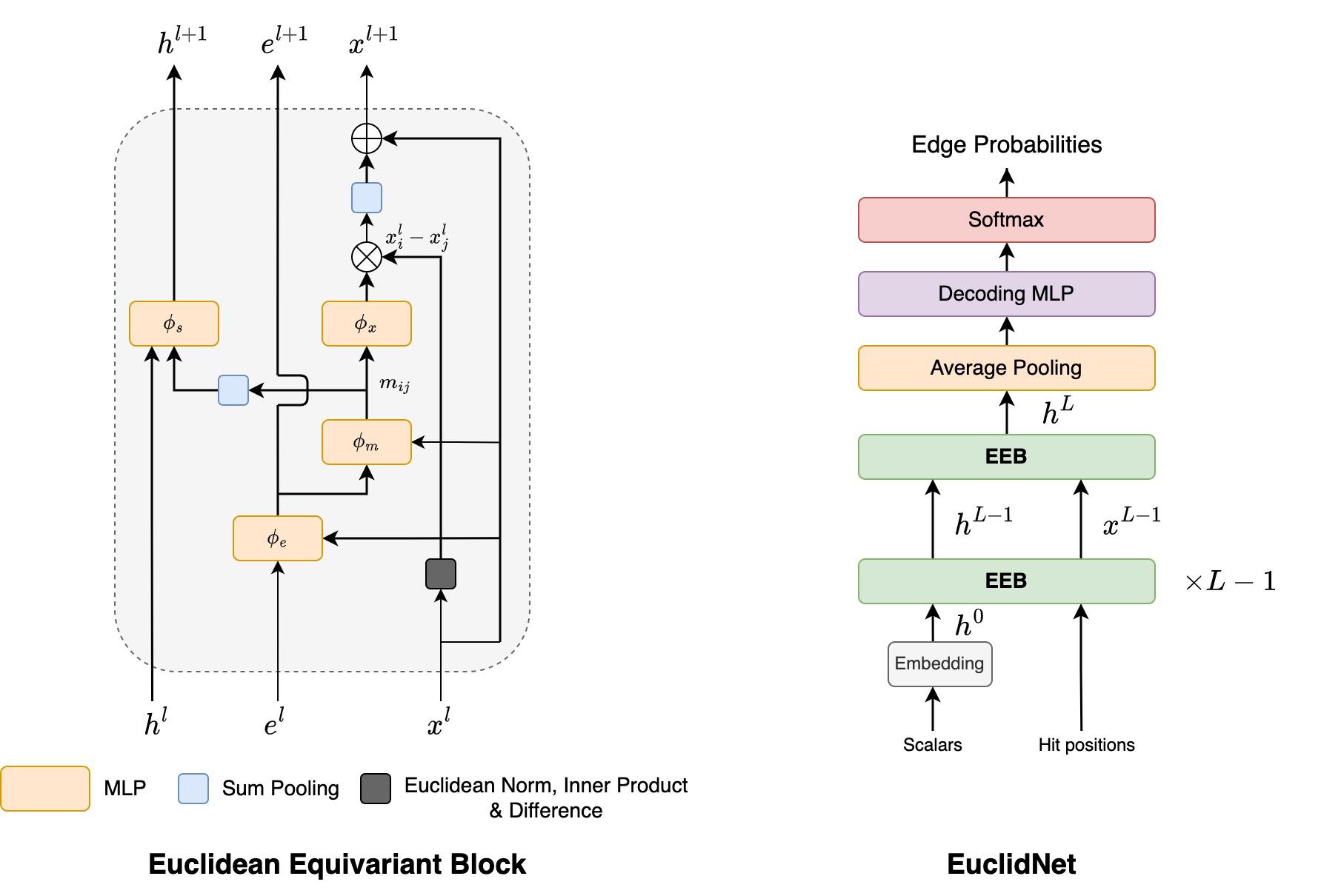}
    \caption{(\textbf{left}) The structure of the Euclidean Equivariant Block (EEB). (\textbf{right}) The network architecture of EuclidNet.}
    \label{fig:euclidnet}
\end{figure}

In this section, we describe the architecture of EuclidNet. As described in Figure \ref{fig:euclidnet}, EuclidNet is constructed by stacking Euclidean Equivariant Blocks (EEB) along with some encoding and decoding layers. The architecture of EuclidNet closely follows that of LorentzNet presented in \cite{Gong:2022lye}. In this section, $\phi(a, b, \dots, f)$ implies that the quantities $a$ through $f$ are concatenated before being passed to $\phi$.\\

\textbf{Input layer.} The inputs to the network are 3D hit positions. For the SO(2) group, scalars are the $z-$coordinate of the hit position, while the 2D coordinates form the vectors. The scalars are projected to an embedding space using an embedding layer before being passed to the first equivariant block.\\

\textbf{Euclidean Equivariant Block.} Following convention, we use $h^l = (h^l_1, h^l_2, \dots, h^l_N)$ to denote node embedding scalars and $x^{l} = (x^l_1, x^l_2, \dots, x^l_N)$ to denote coordinate embedding vectors in the $l-$th EEB layer. $x^0$ corresponds to the hit positions and $h^0$ corresponds to the embedded input of scalar variables. The message $m_{ij}^l$ to be passed is constructed as follows:
\begin{align}
    m^l_{ij} &= \phi_m\left(h^l_i, h^l_j, e^{l}_{ij}, \psi(\Vert {x_i^l - x_j^l} \Vert^2), \psi(\langle x_i^l, x_j^l \rangle)\right)\\
    e_{ij}^{l} &= \phi_e\left(\psi(\Vert {x_i^l - x_j^l} \Vert^2), \psi(\langle x_i^l, x_j^l \rangle), e^{l - 1}_{ij}\right)
\end{align}
where $\phi_m(\cdot)$ is a neural network and $\psi(\cdot) = \rm{sgn}(\cdot)\log(\abs{\cdot} + 1)$ normalizes large numbers from broad distributions to ease training. The input to $\phi_m$ also contains the Euclidean dot product $\langle x_i^l, x_j^l \rangle$. The Euclidean distance $\Vert {x_i^l - x_j^l} \Vert^2 $ between hits is an important feature and we include it for ease of training. $e^{l}_{ij}$ is an edge significance weight learnt by an MLP. 

We use an Euclidean formulation of the dot product attention from \cite{Gong:2022lye} as the aggregation function, which is defined as:
\begin{equation}
    x_i^{l + 1} = x_i^l + c\sum_{j \in \mathcal{N}(i)} \phi_x(m^{l}_{ij}) \cdot (x^l_i - x^l_{j})
\end{equation}
where $\phi_x(\cdot) \in \mathbb{R}$ is a scalar function modeled by an MLP. The hyperparameter $c$ is introduced to control the scale of the updates. The scalar features, $h^{l}_i$, are updated as:
\begin{equation}
    h^{l + 1}_i = h^{l}_i + \phi_h\left( h^{l}_i, \sum_{j \in \mathcal{N}(i)} m^{l}_{ij} \right)
\end{equation}

\textbf{Decoding layer.} After $L$ stacks of EEBs, we decode the node embedding $h^{L} = (h^L_1, h^L_2, \dots, h^L_N)$. Message passing ensures that the information contained in the vector embeddings is propagated to scalars, therefore it is redundant to decode both. A decoding block with three fully connected layers, followed by a softmax function is used to generate a truth score for each track segment.

\vspace{1em}

% The longest section. Can't think of a suitable structure right now. Will try again later.
\section{Results}

\subsection{Dataset}
In this study, we test the developed tracking models on the TrackML dataset, which is a simulated set of proton-proton collision events developed for the TrackML Particle Tracking Challenge \cite{Amrouche:2019wmx}. Events are generated with 200 pileup interactions on average, simulating the high pileup conditions expected at the HL-LHC. Each event contains three-dimensional hit positions and truth information about the particles that generated them. In this work, we limit our discussion to the pixel layers only which consist of a highly-granular set of four barrel and fourteen endcap layers in the innermost region. Each event's tracker hits are converted to a hitgraph through an edge construction algorithm. In addition to transverse momentum, noise, and same-layer filters to modulate the number of hits, graph edges are also required to satisfy constraints on certain geometrical quantities. In this study, we use the geometric graph construction strategy from \cite{DeZoort:2021rbj} to generate graphs, with $p_{T}^{\rm min} = 1.5$ GeV for each event in the dataset. 

\subsection{Experiments}

We train EuclidNet and the Interaction Network for three different values of hidden channels: 8, 16, and 32. The results obtained on the TrackML dataset are summarized in Table \ref{tab:results}. We evaluate the models using the Area Under the ROC curve (AUC), which is a commonly used metric in classification problems. We also report the number of model parameters, as well as the purity (fraction of true edges to total edges in the graph) and efficiency (fraction of all true track segments correctly classified) of the resulting event graphs.  

\begin{table}[!ht]
\caption{Performance comparison between EuclidNet and the Interaction Network (IN) on the TrackML dataset. The results for EuclidNet and IN are averaged over 5 runs}
    \label{tab:results}
    \begin{tabular}{llllll}
        \hline
        \hline
        $N_{hidden}$ & \textbf{Model} & \textbf{Params} & \textbf{AUC} & \textbf{Efficiency} & \textbf{Purity}\\
        \hline
        \multirow{2}{*}{8} & EuclidNet & $967$ & \boldmath{$0.9913 \pm 0.004$} & \boldmath$0.9459 \pm 0.022$ & \boldmath$0.7955 \pm 0.040$\\
        & InteractionNet & 1432 & $0.9849 \pm 0.006$ & $0.9314 \pm 0.021$ & $0.7319 \pm 0.052$\\
        \hline
        \multirow{2}{*}{16} & EuclidNet & $2580$ & $0.9932 \pm 0.003$ & $0.9530 \pm 0.014$ & \boldmath$0.8194 \pm 0.033$\\
        & InteractionNet & 4392 & $0.9932 \pm 0.004$ & \boldmath$0.9575 \pm 0.019$ & $0.8168 \pm 0.073$\\
        \hline
        \multirow{2}{*}{32} & EuclidNet & $4448$ & $0.9941 \pm 0.003$ & $0.9547 \pm 0.019$ & $0.9264 \pm 0.023$\\
        & InteractionNet & 6448 & \boldmath$0.9978 \pm 0.003$ & \boldmath$0.9785 \pm 0.022$ & \boldmath$0.9945 \pm 0.043$\\
        \hline
    \end{tabular}\hspace*{-64pt}
\end{table}

\begin{figure}
    \centering
    \begin{subfigure}[b]{0.46\textwidth}
    \includegraphics[width=\textwidth]{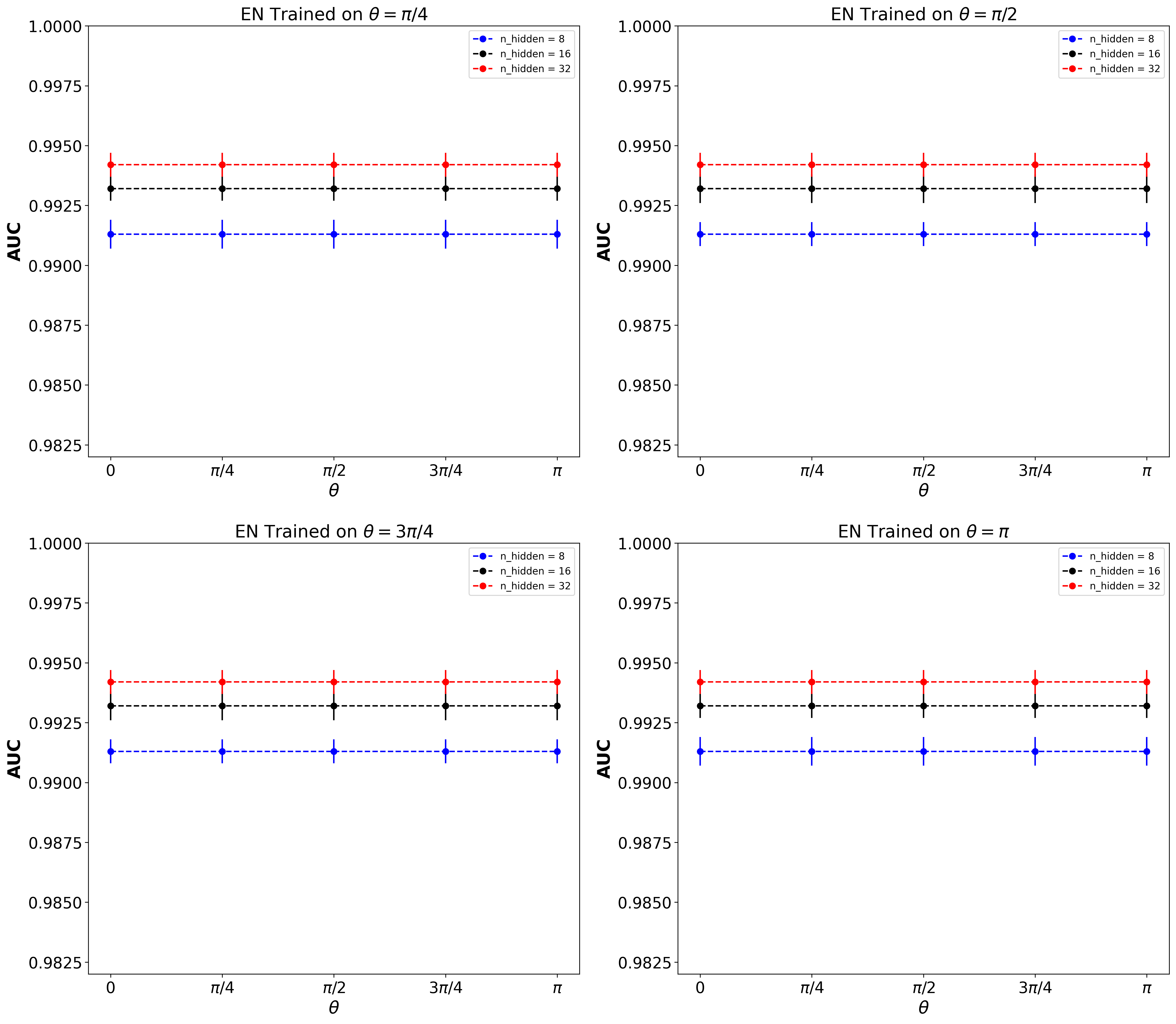}
    \label{fig:image1}
  \end{subfigure}
  \hspace{1cm}
  \begin{subfigure}[b]{0.46\textwidth}
    \includegraphics[width=\textwidth]{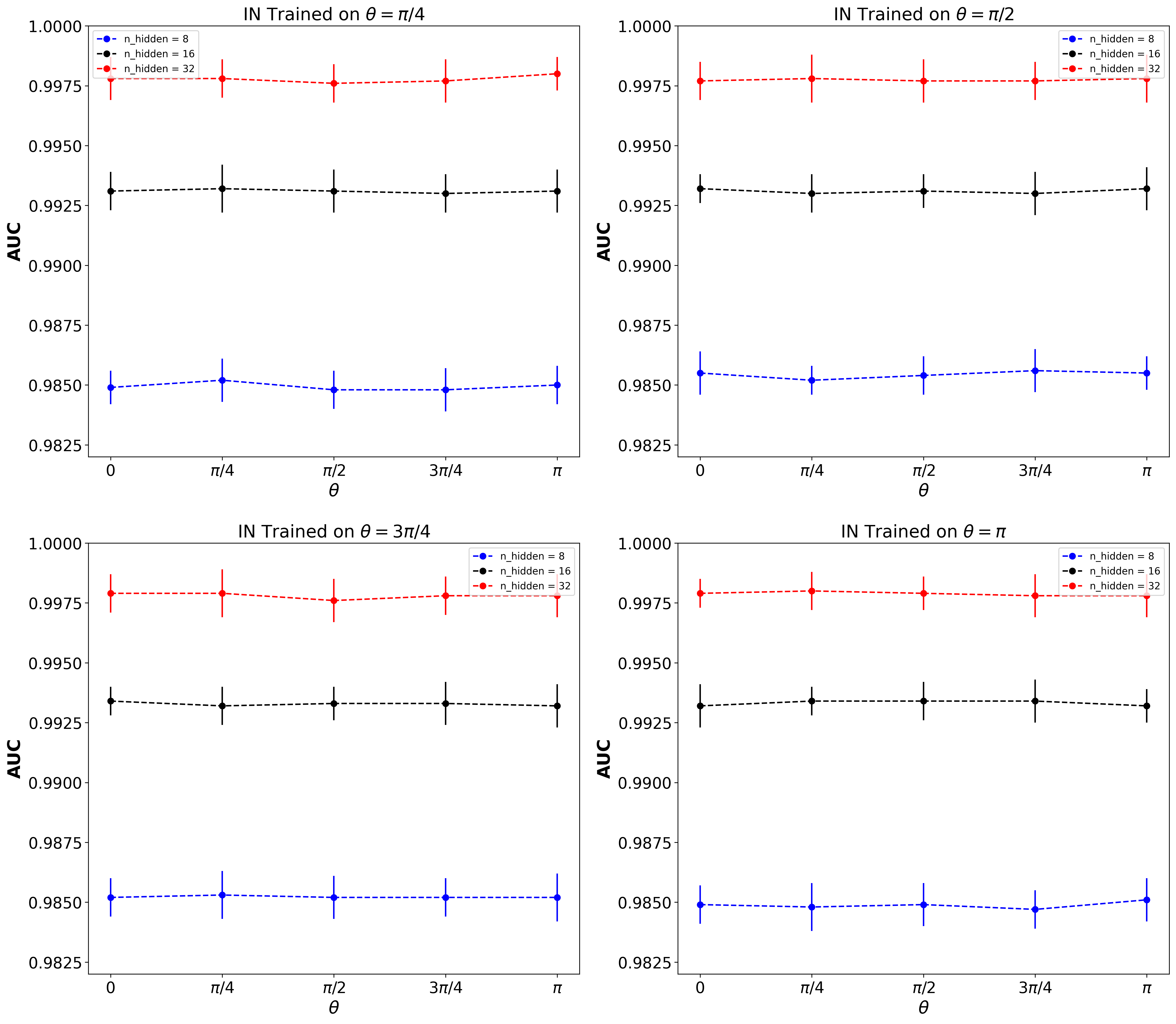}
    \label{fig:image2}
  \end{subfigure}
  \caption{The AUC at inference time plotted as a function of rotations in the input space by an angle $\theta$ for models trained on a set of rotations of the dataset by $\theta \in [0, \pi/4, \pi/2, 3\pi/4, \pi]$ for (\textbf{left}) EuclidNet and (\textbf{right}) the Interaction Network calculated over 5 independent inference runs.}
  \label{fig:rotation_tests}
\end{figure}

For small model sizes, EuclidNet outperforms the unconstrained Interaction Network. Rotational symmetry is approximately obeyed in the generic TrackML detector and using this symmetry appears, to first order, to produce an accurate edge classification GNN. However, given a larger latent space, the performance with rotational symmetry enforced plateaus. At that point, the unconstrained network is more performant. We hypothesize that, as in real detectors like ATLAS and CMS, the rotational symmetry of the TrackML dataset is likely only approximate due to the granularity of the detector segments and inhomogeneity in the magnetic field surrounding the detector. To study this possibility, we train both models on a set of rotations of the dataset by $\theta \in [0, \pi/4, \pi/2, 3\pi/4, \pi]$, and run inference of each of these instances across the set of rotated datasets. In this way, we can capture whether the unconstrained network is learning a function specific to that orientation of the detector's material and magnetic field. The results of this study are summarized in Figure \ref{fig:rotation_tests}.

In Figure \ref{fig:rotation_tests}, we observe that both the SO(2)-equivariant EuclidNet and the Interaction Network are robust to rotations in input space, irrespective of which set of rotations of the dataset the model was trained on. The general trend observed in Table \ref{tab:results} is also replicated here, with models having a larger latent space producing larger AUC scores. Although further study is needed to interpret the unconstrained network's superior performance, we posit that this might be a artifact of the dataset's inherent approximate rotational symmetry. In such case, then a less expressive EuclidNet would not be able to completely match the approximate symmetry of the event, and would outperform the IN only at latent space sizes that are not sufficient for IN to capture the complete symmetry set. However, given a large enough latent space, an unconstrained network such as the IN appears to easily learn both the approximate symmetry and non-symmetric corrections in the dataset. In the future, an more in-depth study leveraging explainable AI techniques and methods like `LieGG' \cite{moskalev2023liegg} to determine the actual symmetry learnt by both EuclidNet and the IN is required to ascertain the cause for this behaviour.  

\vspace{0.5em}

\section{Conclusions and Future Work}

In this study, we have presented EuclidNet --- the first Euclidean rotation-equivariant GNN for the particle tracking problem. The SO(2)-equivariant model offers a marginal improvement (AUC = 0.9913) over the benchmark (AUC = 0.9849) at small model scales ($ < 1000$ parameters). However, for the particle tracking problem, we find that an unconstrained model still outperforms an equivariant architecture at larger model sizes. More work is needed to concretely establish the reasons for this result. Possible future directions include studying the problem’s equivariance and identifying non-equivariant facets, if they exist, as well as investigating the quality of the learnt symmetry. However, if the dataset inherently contains non-equivariant features, any symmetric model will likely always underperform. In this case, models such as in \cite{Murnane_2023} which relax the strict constraints imposed by symmetry-following architectures might be able to learn the non-equivariant aspects of the tracking dataset.

\section*{Acknowledgements}
This work was supported by IRIS-HEP through the U.S. National Science Foundation under Cooperative Agreement OAC-1836650. This research used resources of the National Energy Research Scientific Computing Center (NERSC), a U.S. Department of Energy Office of Science User Facility located at Lawrence Berkeley National Laboratory, operated under Contract No. DE-AC02-05CH11231.

\appendix
\section{Training Details}
We use the binary cross-entropy loss function for both EuclidNet and the Interaction Network. All models are optimized using the Adam optimizer \cite{Kingma:2014vow} implemented in PyTorch with a learning rate $\eta = 1 \times 10^{-3}$ and momentum coefficients $(\beta_1, \beta_2) = (0.9, 0.999)$. The learning rate is decayed by $0.3$ every 100 epochs. Both are trained on a single NVIDIA A100 GPU each for a 200 epochs using early stopping with a patience of 50 epochs. Total training time for both models is typically 2 hours. 

\section{Equivariance Tests}
\begin{figure}[!h]
    \centering
    \includegraphics[scale=0.45]{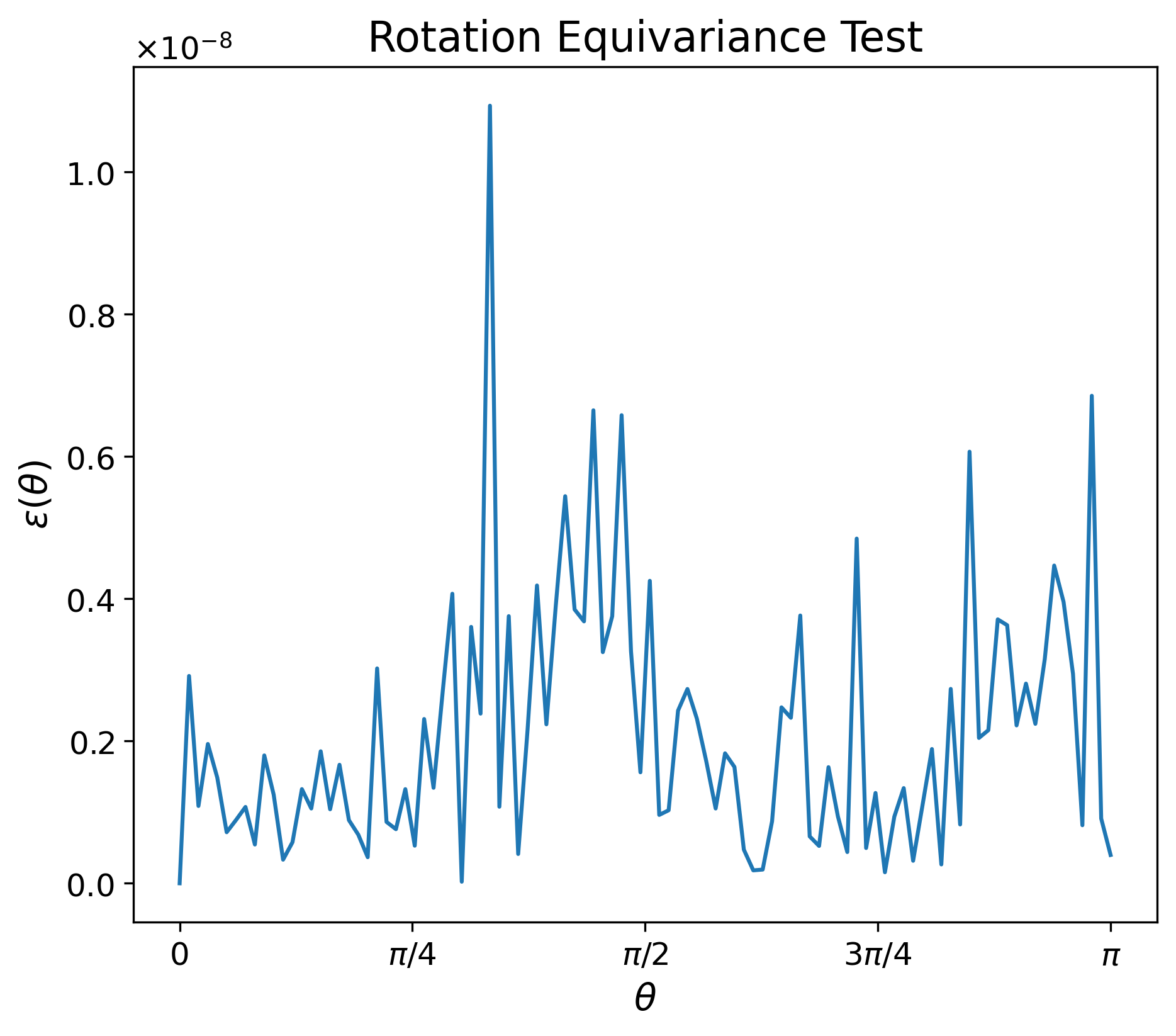}
    \caption{Relative deviation of the output node vector representations to rotations around the beamline ($z$-axis).}
    \label{fig:eqv-test}
\end{figure}

We also test the equivariance of EuclidNet and find that it is indeed equivariant to SO(2) transformations up to floating-point numerical errors. For a given transformation $R \in SO(2)$, we compare $R_{\theta}\phi(\vb{x})$ and $\phi(R_{\theta}\vb{x})$, where $\vb{x}$ are the hit coordinates and $\phi(\cdot)$ is a EuclidNet instance. In Figure \ref{fig:eqv-test} we plot the relative deviation, defined as
\begin{equation}
    \varepsilon(\theta) = \frac{\langle \phi(R_{\theta}\vb{x}) \rangle - \langle R_{\theta}\phi(\vb{x}) \rangle}{\langle R_{\theta}\phi(\vb{x}) \rangle}
\end{equation}
where $\langle \cdot \rangle$ is the mean computed over 500 tracking events in the test dataset. We find the relative deviation from rotations to be $< 10^{-8}$. 

\bibliography{bibliography}

\end{document}